\documentclass[12pt]{article}
\usepackage{amsmath,amssymb,amsthm}
\usepackage{graphics,epsfig}
\usepackage{hyperref}
\usepackage{natbib}
\usepackage{color}
\usepackage{graphicx}
\usepackage{caption}
\usepackage{subcaption}
\usepackage{float}
\usepackage{mathrsfs}
\usepackage{multirow}
\usepackage{comment}
\usepackage{setspace}
\usepackage{bbm}
\usepackage{bm}
\usepackage{amsfonts}
\usepackage{xcolor}
\usepackage{pslatex}
\usepackage{setspace}
\usepackage{bbm}

\def \E {\mathbb{E}}

\def \a {\alpha}


\begin{document}

  \title{\bfseries Feedback strategies in the markets with uncertainties}

  \author{
Mustapha Nyenye Issah\footnote{Corresponding author, {\small\texttt{mi2223@jagmail.southalabama.edu}}}\; \footnote{Department of Mathematics and Statistics,  University of South Alabama, Mobile, AL, 36688,
United States.}
}

\date{\today}
\maketitle

\begin{abstract}
We explore how dynamic entry deterrence operates through feedback strategies in markets experiencing stochastic demand fluctuations. The incumbent firm, aware of its own cost structure, can deter a potential competitor by strategically adjusting prices. The potential entrant faces a one-time, irreversible decision to enter the market, incurring a fixed cost, with profits determined by market conditions and the incumbent’s hidden type. Market demand follows a Chan-Karolyi-Longstaff-Sanders (CKLS) Brownian motion. If the demand is low, the threat of entry diminishes, making deterrence less advantageous. In equilibrium, a weak incumbent may be incentivized to reveal its type by raising prices. We derive an optimal equilibrium using path integral control, where the entrant enters once demand reaches a high enough level, and the weak incumbent mixes strategies between revealing itself when demand is sufficiently low.
\end{abstract}

\section*{Introduction:}
Companies strategically set prices to influence the entry and exit of rivals, with limit pricing being a notable tactic where an incumbent keeps prices low to discourage potential competitors. The airline industry showcases clear examples of this behavior, and research has shown that ticket prices tend to drop when competition looms \citep{pramanik2020optimization,gryglewicz2023strategic}. In reality, such strategic pricing occurs in dynamic and uncertain environments, requiring incumbents to signal unprofitable entry through repeated pricing decisions amid fluctuating market conditions. Since strategic pricing involves balancing short-term costs against long-term gains, the decision to employ it hinges on evolving market demand. Likewise, the potential entrant's decision to enter is influenced by these shifting market conditions. This paper aims to explore the dynamics of these interactions in a context where market demand experiences persistent shocks \citep{gryglewicz2023strategic}.

We analyze a two-player, continuous-time framework where a market incumbent has private knowledge of its marginal costs, classifying it as either strong or weak. The potential entrant is uncertain about the incumbent’s cost structure and can choose to enter the market by incurring a fixed cost \citep{pramanik2024dependence}. Upon entry, the entrant receives a positive, increasing payoff in line with market demand if the incumbent is weak, but earns nothing if the incumbent is strong. A weak incumbent faces a lower payoff after entry but can deter it by engaging in a costly signaling action, which can be understood as setting low prices to imitate a strong incumbent. We model market demand as a Chan-Karolyi-Longstaff-Sanders (CKLS) family of Brownian motions since this captures the dynamics of a time series that exhibits mean-reversion and volatility that changes with the level of the variable (i.e., heteroskedasticity) \citep{chan1992empirical,pramanik2021optimal,pramanik2023scoring}. This family was originally developed to improve models like the Vasicek or Cox-Ingersoll-Ross (CIR) for state variables by allowing more flexibility in how volatility behaves in response to changes in state variables. It accounts for both the mean-reversion of interest rates and allows the volatility of interest rates to vary with the level of the rate, which is crucial in financial markets where interest rate changes do not behave uniformly \citep{pramanik2021optimization}. Furthermore, if markets are subject to stochastic demand shocks, the CKLS family can model how these shocks affect state and control variables over time, especially when volatility and mean-reversion are relevant. This is applicable in sectors such as energy markets, commodity pricing, and in sports \citep{pramanik2020motivation}. Finally, in strategic business situations, such as dynamic entry deterrence models, CKLS can represent market demand or cost structures that evolve stochastically \cite{pramanik2024optimization}. For example, in markets where incumbents and potential entrants react to changing demand conditions, the CKLS model helps in forecasting market conditions under different volatility regimes. We permit this process to include either upward or downward drift, making this model applicable to markets that are generally growing or shrinking \citep{pramanik2024dependence}.

How does the incumbent signal in such a dynamic environment? How does the entrant strategically time its market entry? Our key finding is that in markets experiencing persistent shocks, the incumbent’s incentive to signal vanishes under certain conditions \citep{pramanik2024semicooperation}. The entrant aims to enter when demand is sufficiently high compared to its revised belief that the incumbent is strong. Conversely, when demand drops low enough, the threat of entry diminishes in the short term, leading a weak incumbent to forgo costly signaling and instead reveal its type by increasing prices \citep{carmona2015forward}. 

We propose an optimal feed-back Nash equilibrium of strategies between the incumbent and entrant. One strategy might be the total amount of expenditure on promotions of commercials. This is a type of closed loop strategy (i.e.,control) since the optimal strategy is determined by the current market size $X\in\mathcal X$ and time $s$. We define firm $i^{th}$ strategy as $u^i(s):=\left\{\tilde{u}^i(s,X)|0\leq s\leq t,X\in\mathcal X\right\}$, for all $i=1,2$. In feed-back strategy the decisions might be revised based on new information embodied in current market size. We propose a Feynman-type path integral control method to determine these strategies \citep{pramanik2021consensus,pramanik2023optimal,pramanik2023path1,pramanik2023path}. The benefit of this approach is that it bypasses the complex calculation of the value function. However, when the market size is extremely large, solving the Hamiltonian-Jacobi-Bellman (HJB) equation numerically becomes highly challenging \citep{pramanik2022lock,pramanik2022stochastic}. As a result, method like the Feynman-type path integral approach is useful. Additionally, if the market dynamics are nonlinear, finding an optimal equilibrium via the HJB equation is not feasible. In such cases, the Feynman-type path integral method proves effective \citep{pramanik2024parametric}.

\section*{Construction of the model:}
The game take place in continuous time $s\in[0,t]$ over a finite horizon. Firm 1 possesses a private characteristic (either weak or strong) and seeks to persuade firm 2 that it is strong. Both firms confront optimal stopping problems as market demand evolves publicly: firm 1 (if weak) decides the timing of disclosing its type, while firm 2 determines the optimal moment to enter the market. Consider the background  probability space of the market size is $(\Omega,\mathcal F,\mathcal P)$, where $\Omega$ is the sample space, $\mathcal F$ is the Borel-$\sigma$ field, and $\mathcal P$ is the probability measure. The construction of the probability space is the following: first, there exists a standard Brownian motion $B(s)$ defined on the canonical probability space $(\tilde\Omega,{\tilde{\mathcal F}},{\tilde{\mathcal P}})$. The incumbent has the probability space $(\Theta,2^{\Theta},\eta)$, where $\Theta={q,w}$ such that if incumbent is strong (i.e., $q$) or weak (i.e., $w$). The entrant's prior is $\eta(\theta=Q)=p_0\in[0,1]$. Therefore, in this construction $\Omega=\tilde\Omega\times\Theta$, $\mathcal F={\tilde{\mathcal F}}\times 2^\Theta$, and $\mathcal P={\tilde{\mathcal P}}\times \eta$. The dynamics of market size evolves over time interval $[0,t]$, and can be expressed by the following CKLS stochastic differential equation (SDE) \citep{yeung2006cooperative}
\begin{equation}\label{0}
dX(s)=\left[\a_1+\a_2 X(s)+\sum_{i=1}^2\theta_i u^i(s)\right]ds+\a_3[X(s)]^{\a_4}dB(s),
\end{equation} 
where $\a_j$ for all $j=1,\dots,4$ and $\theta_i$, for every $i=1,2$ are constants, $\left[\a_1+\a_2 X(s)+\sum_{i=1}^2\theta_i u^i(s)\right]$ is the drift coefficient, $\a_4[X(s)]^{\a_5}$ is the diffusion coefficient, and $B(s)$ is a two-dimensional Brownian motion in the probability space $(\Omega^X,\mathcal F^X,\mathcal P^X)$ such that $\mathcal F^X=\left\{\mathcal F_s^X\right\}_{s\in[0,t]}$, where $\mathcal F_s^X=\sigma\left(X(\nu),Y_0|\nu\in[0,\tau]\right)$ be the natural filtration of $X$, and $X(0)=x_0>0$ be the initial market size \citep{polansky2021motif,pramanik2024estimation}. To allow this random mechanism, an independent random device $Y_0$ observed at $s=0$. A history $h(s)$ is the realization of $Y_0$ and the sample path $\omega$ of $X$ from time $0$ to $t$. Both firms update their beliefs of $p(s)$ based on Bayes' rule \citep{mcclellan2022experimentation}.

The entrant and the weak incumbent face interconnected stopping problems, while the strong incumbent is considered non-strategic and does not engage in explicitly modeled behavior. The weak incumbent selects a time $\rho$ to disclose its type publicly, and the entrant decides on a time $\tilde\tau$ to enter the market \citep{pramanik2023path1,pramanik2023path}. It is assumed that the entrant observes the incumbent's decision at time $s$ before making its own move, but the incumbent cannot observe the entrant's decision in return \citep{pramanik2024measuring}. Upon entry, the entrant pays a fixed entry fee $F>0$, which results in the incumbent's type being revealed. Following this, if the incumbent is weak, the entrant receives a duopoly flow payoff of $D_E^w X(s)$,  and nothing if the incumbent is strong. Meanwhile, the weak incumbent earns a duopoly flow payoff of  $D_I^w X(s)$. These post-entry flow payoffs lead to termination payoffs for the pre-entry stopping game. Before entering, the entrant earns no flow payoff, while the weak incumbent receives a signaling flow payoff of $QX(s)$ until it reveals its type, followed by a monopoly flow payoff of $M^wX(s)$ \citep{hua2019assessing,pramanik2016tail,pramanik2021effects}. 

To keep the game engaging yet manageable, we introduce the following parametric assumptions. First, entering is never advantageous against type $q$, but can be profitable against type $w$ when the market size is high enough to cover the fixed costs: $D^W_E > 0$. Second, for type $w$, continuing to signal incurs a cost: $M^w > Q$. Third, type $w$ favors signaling over dealing with immediate entry: $Q > D^W_E$. We further assume, both the firms are risk neutral and discounts payoff at a constant rate $\gamma$ such that 
\[
\a_1+\a_2 X(s)+\sum_{i=1}^2\theta_i u^i(s)\leq \gamma
\]
to ensure a finite discounted payoffs for both of them. At time $s$, both firms revise their beliefs to $p(s)$ based on Bayes' rule. For simplicity, we refer to the log-likelihood of these beliefs as $Z(s) = \ln\left[\frac{p(s)}{1-p(s)}\right]$, which we will simply call ``beliefs" \citep{mcclellan2022experimentation}. The belief at $s$ can be expressed as 
\begin{equation}\label{1}
Z(s)=Z_0+\frac{2}{\left(\a_4\right)^2[X(s)]^{2\a_5}}\left[\a_1+\a_2 X(s)+\sum_{i=1}^2\theta_i u^i(s)\right]\left[X(s)-x_0\right].
\end{equation}
Furthermore, we define $z_0:=\ln\left[\frac{p(0)}{1-p(0)}\right]$. The main logic behind this definition is that both $X(s)$ and initial belief $Z_0$ enter linearly in Equation \eqref{1}. We denote $\E_{0,z_0}\{.\}$ the expectation over $X$, $Y_0$ so that $X(0)=x$ and the beliefs $Z_0=z_0$. The expected payoff of firm 1 is 
\begin{multline}\label{2}
J^1\left(\rho,t,x_0,z_0,u^1\right)=\E_{0,z_0}\biggr\{\int_0^{\rho\wedge t}e^{-\gamma s}[Q-u^1(s)]X(s)ds+\mathbbm{1}(\rho\leq t)\int_\rho^te^{-\gamma s}[M^W-\bar u^3]X(s)ds\\+e^{-\gamma t}\frac{D_I^WX(t)}{\gamma-\left[\a_1+\a_2 X(t)+\sum_{i=1}^2\theta_i u^i(t)\right]}\biggr\},
\end{multline}
where $\mathbbm{1}(\rho\leq t)$ is a simple function which takes the value $1$ if $\rho\leq t$ and zero otherwise. Given strategy $u^1(s)$ of incumbent, entrant's expected payoff is 
\begin{equation}\label{3}
J^2\left(t,x_0,z_0,u^1\right)=\E_{0,z_0}\left\{e^{-\gamma t}\left[\mathbbm{1}(\theta=w)\frac{\left[D_E^W-[u^2(t)]^2\right]X(t)}{\gamma-\left[\a_1+\a_2 X(t)+\sum_{i=1}^2\theta_i u^i(t)\right]}-F\right]\right\}.
\end{equation}

In this study, we examine a feedback control methodology. The stochastic control literature primarily distinguishes between two types of control: open-loop and closed-loop. In an open-loop system, the optimal control path is expressed as a function of time, denoted by $u^i(s)$. This control is fully defined at the initial time $s=0$, with the state trajectory $X(s)$ derived by solving the system of SDEs starting from the initial conditions \citep{intriligator2002mathematical}. Conversely, in a closed-loop system, the optimal control path depends on both the current state variables and time, represented as $u^i\left[s,X(s)\right]$. Unlike open-loop control, where decisions are fixed in advance, closed-loop control allows for adjustments based on updated information reflected in the current state variables. 

Automatic stabilizers like unemployment benefits and progressive income taxes operate as closed-loop systems. When unemployment rises, government payments through unemployment insurance increase, mitigating the impact of rising unemployment. Likewise, with higher inflation, the progressive tax system imposes proportionally higher taxes, helping to offset inflationary pressures. In both instances, the control mechanisms adjust in response to the economy’s current state. Another closed-loop system in the economy is the Federal Reserve's monetary policy, which adjusts money and credit supply based on prevailing economic conditions. However, some have proposed transforming this into an open-loop system, where a predetermined rate of money supply growth, such as five percent annually, would be implemented irrespective of current economic trends \citep{intriligator2002mathematical}. Closed-loop control typically outperforms open-loop control when it comes to maximizing the pay-off function, particularly in stochastic control scenarios, where random variables with known distributions are involved. It also excels in feedback control situations, where initial uncertainties regarding the parameters, functions, or constraints of the problem are gradually reduced or resolved as the process progresses. It is important to note that, feedback control is indeed a closed loop control.

The interaction between the incumbent and the entrant follows a Stackelberg-like framework, where the incumbent acts first by selecting its output before the entrant. The entrant observes this decision and then determines its own output, taking the incumbent's choice into account. We refer to this as Stackelberg-like because the entrant is initially unaware of the incumbent's monopoly power. At time $\rho$, if the entrant realizes that the incumbent lacks monopoly power, it enters the market and generates profit over the continuous time period $[\rho,t]$ \citep{pramanik2020motivation}. Otherwise, the entrant chooses not to enter the market. Therefore, the incumbent maximizes \eqref{2} subject to Equations \eqref{3} and \eqref{0}. On the other hand, if the entrant decides to enter the market at time $\rho$, it maximizes

\begin{equation}\label{4}
J^2\left(t,x_0,z_0,u^1\right)=\E_{0,z_0}\left\{\int_\rho^s e^{-\gamma s}\left[\mathbbm{1}(\theta=w)\frac{\left[D_E^W-[u^2(s)]^2\right]X(s)}{\gamma-\left[\a_1+\a_2 X(s)+\theta_1 u^{1*}(s)+\theta_2u^2(s)\right]}\right]ds-F\right\},
\end{equation}
subject to
\begin{equation}\label{5}
dX(s)=\left[\a_1+\a_2 X(s)+\theta_1 u^{1*}(s)+\theta_2u^2(s)\right]ds+\a_3[X(s)]^{\a_4}dB(s),
\end{equation}
 where $u^{1*}(s)$ is the optimal promotional expenditure on commercials which is observed by the entrant at $\rho$. After performing a Feynman-type path integral control approach we obtain a feedback Nash equilibrium $\left\{u^{1*}(s),u^{2*}(s)\right\}$. To apply this method, we begin by formulating a stochastic Lagrangian at each point in continuous time within the interval $s \in [0, t]$, where $t > 0$. Next, we partition the entire time period into $n$ subintervals of equal length and assign a Riemann measure for the state variable in each subinterval \citep{pramanik2024motivation}. After constructing a Euclidean action function, we derive a Schr\"odinger-type equation using a Wick rotation. The system's solution is then determined by applying the first-order conditions with respect to both the state and control variables. This approach is fruitful in cancer research \citep{dasgupta2023frequent,hertweck2023clinicopathological,kakkat2023cardiovascular,khan2023myb,vikramdeo2023profiling,khan2024mp60,vikramdeo2024abstract}.
 
 \section*{Discussion:}
In this paper we discuss a  model of entry deterrence through strategic pricing in a stochastic setting, demonstrating that this framework might create an optimal strategic interactions between the incumbent and the entrant \citep{pramanik2023optimal}. Based on their believes about the incumbent's type, the entrant is motivated to enter when the stakes are high enough relative to the incumbent's reputation \citep{pramanik2021consensus,pramanik2024bayes}. Conversely, the incumbent is inclined to stop signaling when the stakes are sufficiently low, and in equilibrium, a weaker type of incumbent separates itself gradually through randomization. Even though our model does not involve any exogenous information about the incumbent, it still generates complex belief dynamics that are driven by the running minimum of the stakes. Under this circumstance an optimal can be achieved through a path integral approach.

\bibliographystyle{apalike}
\bibliography{bib}

\begin{thebibliography}{}

\bibitem[Carmona and Delarue, 2015]{carmona2015forward}
Carmona, R. and Delarue, F. (2015).
\newblock Forward--backward stochastic differential equations and controlled
  mckean--vlasov dynamics.
\newblock {\em The Annals of Probability}, pages 2647--2700.

\bibitem[Chan et~al., 1992]{chan1992empirical}
Chan, K.~C., Karolyi, G.~A., Longstaff, F.~A., and Sanders, A.~B. (1992).
\newblock An empirical comparison of alternative models of the short-term
  interest rate.
\newblock {\em The journal of finance}, 47(3):1209--1227.

\bibitem[Dasgupta et~al., 2023]{dasgupta2023frequent}
Dasgupta, S., Acharya, S., Khan, M.~A., Pramanik, P., Marbut, S.~M., Yunus, F.,
  Galeas, J.~N., Singh, S., Singh, A.~P., and Dasgupta, S. (2023).
\newblock Frequent loss of cacna1c, a calcium voltage-gated channel subunit is
  associated with lung adenocarcinoma progression and poor prognosis.
\newblock {\em Cancer Research}, 83(7\_Supplement):3318--3318.

\bibitem[Gryglewicz and Kolb, 2023]{gryglewicz2023strategic}
Gryglewicz, S. and Kolb, A. (2023).
\newblock Strategic pricing in volatile markets.
\newblock {\em Operations Research}.

\bibitem[Hertweck et~al., 2023]{hertweck2023clinicopathological}
Hertweck, K.~L., Vikramdeo, K.~S., Galeas, J.~N., Marbut, S.~M., Pramanik, P.,
  Yunus, F., Singh, S., Singh, A.~P., and Dasgupta, S. (2023).
\newblock Clinicopathological significance of unraveling mitochondrial pathway
  alterations in non-small-cell lung cancer.
\newblock {\em The FASEB Journal}, 37(7):e23018.

\bibitem[Hua et~al., 2019]{hua2019assessing}
Hua, L., Polansky, A., and Pramanik, P. (2019).
\newblock Assessing bivariate tail non-exchangeable dependence.
\newblock {\em Statistics \& Probability Letters}, 155:108556.

\bibitem[Intriligator, 2002]{intriligator2002mathematical}
Intriligator, M.~D. (2002).
\newblock {\em Mathematical optimization and economic theory}.
\newblock SIAM.

\bibitem[Kakkat et~al., 2023]{kakkat2023cardiovascular}
Kakkat, S., Pramanik, P., Singh, S., Singh, A.~P., Sarkar, C., and Chakroborty,
  D. (2023).
\newblock Cardiovascular complications in patients with prostate cancer:
  Potential molecular connections.
\newblock {\em International Journal of Molecular Sciences}, 24(8):6984.

\bibitem[Khan et~al., 2023]{khan2023myb}
Khan, M.~A., Acharya, S., Anand, S., Sameeta, F., Pramanik, P., Keel, C.,
  Singh, S., Carter, J.~E., Dasgupta, S., and Singh, A.~P. (2023).
\newblock Myb exhibits racially disparate expression, clinicopathologic
  association, and predictive potential for biochemical recurrence in prostate
  cancer.
\newblock {\em Iscience}, 26(12).

\bibitem[Khan et~al., 2024]{khan2024mp60}
Khan, M.~A., Acharya, S., Kreitz, N., Anand, S., Sameeta, F., Pramanik, P.,
  Keel, C., Singh, S., Carter, J., Dasgupta, S., et~al. (2024).
\newblock Mp60-05 myb exhibits racially disparate expression and
  clinicopathologic association and is a promising predictor of biochemical
  recurrence in prostate cancer.
\newblock {\em The Journal of Urology}, 211(5S):e1000.

\bibitem[McClellan, 2022]{mcclellan2022experimentation}
McClellan, A. (2022).
\newblock Experimentation and approval mechanisms.
\newblock {\em Econometrica}, 90(5):2215--2247.

\bibitem[Polansky and Pramanik, 2021]{polansky2021motif}
Polansky, A.~M. and Pramanik, P. (2021).
\newblock A motif building process for simulating random networks.
\newblock {\em Computational Statistics \& Data Analysis}, 162:107263.

\bibitem[Pramanik, 2016]{pramanik2016tail}
Pramanik, P. (2016).
\newblock {\em Tail non-exchangeability}.
\newblock Northern Illinois University.

\bibitem[Pramanik, 2020]{pramanik2020optimization}
Pramanik, P. (2020).
\newblock Optimization of market stochastic dynamics.
\newblock In {\em SN Operations Research Forum}, volume~1, page~31. Springer.

\bibitem[Pramanik, 2021a]{pramanik2021consensus}
Pramanik, P. (2021a).
\newblock Consensus as a nash equilibrium of a stochastic differential game.
\newblock {\em arXiv preprint arXiv:2107.05183}.

\bibitem[Pramanik, 2021b]{pramanik2021effects}
Pramanik, P. (2021b).
\newblock Effects of water currents on fish migration through a feynman-type
  path integral approach under 8/3 liouville-like quantum gravity surfaces.
\newblock {\em Theory in Biosciences}, 140(2):205--223.

\bibitem[Pramanik, 2021c]{pramanik2021optimization}
Pramanik, P. (2021c).
\newblock {\em Optimization of dynamic objective functions using path
  integrals}.
\newblock PhD thesis, Northern Illinois University.

\bibitem[Pramanik, 2022a]{pramanik2022lock}
Pramanik, P. (2022a).
\newblock On lock-down control of a pandemic model.
\newblock {\em arXiv preprint arXiv:2206.04248}.

\bibitem[Pramanik, 2022b]{pramanik2022stochastic}
Pramanik, P. (2022b).
\newblock Stochastic control of a sir model with non-linear incidence rate
  through euclidean path integral.
\newblock {\em arXiv preprint arXiv:2209.13733}.

\bibitem[Pramanik, 2023a]{pramanik2023optimal}
Pramanik, P. (2023a).
\newblock Optimal lock-down intensity: A stochastic pandemic control approach
  of path integral.
\newblock {\em Computational and Mathematical Biophysics}, 11(1):20230110.

\bibitem[Pramanik, 2023b]{pramanik2023path1}
Pramanik, P. (2023b).
\newblock Path integral control in infectious disease modeling.
\newblock {\em arXiv preprint arXiv:2311.02113}.

\bibitem[Pramanik, 2023c]{pramanik2023path}
Pramanik, P. (2023c).
\newblock Path integral control of a stochastic multi-risk sir pandemic model.
\newblock {\em Theory in Biosciences}, 142(2):107--142.

\bibitem[Pramanik, 2024a]{pramanik2024dependence}
Pramanik, P. (2024a).
\newblock Dependence on tail copula.
\newblock {\em J}, 7(2):127--152.

\bibitem[Pramanik, 2024b]{pramanik2024estimation}
Pramanik, P. (2024b).
\newblock Estimation of optimal lock-down and vaccination rate of a stochastic
  sir model: A mathematical approach.
\newblock {\em European Journal of Statistics}, 4:3--3.

\bibitem[Pramanik, 2024c]{pramanik2024measuring}
Pramanik, P. (2024c).
\newblock Measuring asymmetric tails under copula distributions.
\newblock {\em European Journal of Statistics}, 4:7--7.

\bibitem[Pramanik et~al., 2024]{pramanik2024parametric}
Pramanik, P., Boone, E.~L., and Ghanam, R.~A. (2024).
\newblock Parametric estimation in fractional stochastic differential equation.
\newblock {\em Stats}, 7(3):745.

\bibitem[Pramanik and Maity, 2024]{pramanik2024bayes}
Pramanik, P. and Maity, A.~K. (2024).
\newblock Bayes factor of zero inflated models under jeffereys prior.
\newblock {\em arXiv preprint arXiv:2401.03649}.

\bibitem[Pramanik and Polansky, 2020]{pramanik2020motivation}
Pramanik, P. and Polansky, A.~M. (2020).
\newblock Motivation to run in one-day cricket.
\newblock {\em arXiv preprint arXiv:2001.11099}.

\bibitem[Pramanik and Polansky, 2021]{pramanik2021optimal}
Pramanik, P. and Polansky, A.~M. (2021).
\newblock Optimal estimation of brownian penalized regression coefficients.
\newblock {\em arXiv preprint arXiv:2107.02291}.

\bibitem[Pramanik and Polansky, 2023]{pramanik2023scoring}
Pramanik, P. and Polansky, A.~M. (2023).
\newblock Scoring a goal optimally in a soccer game under liouville-like
  quantum gravity action.
\newblock In {\em Operations Research Forum}, volume~4, page~66. Springer.

\bibitem[Pramanik and Polansky, 2024a]{pramanik2024motivation}
Pramanik, P. and Polansky, A.~M. (2024a).
\newblock Motivation to run in one-day cricket.
\newblock {\em Mathematics}, 12(17):2739.

\bibitem[Pramanik and Polansky, 2024b]{pramanik2024optimization}
Pramanik, P. and Polansky, A.~M. (2024b).
\newblock Optimization of a dynamic profit function using euclidean path
  integral.
\newblock {\em SN Business \& Economics}, 4(1):8.

\bibitem[Pramanik and Polansky, 2024c]{pramanik2024semicooperation}
Pramanik, P. and Polansky, A.~M. (2024c).
\newblock Semicooperation under curved strategy spacetime.
\newblock {\em The Journal of Mathematical Sociology}, 48(2):172--206.

\bibitem[Vikramdeo et~al., 2024]{vikramdeo2024abstract}
Vikramdeo, K., Anand, S., Sudan, S., Pramanik, P., Singh, S., Godwin, A.,
  Singh, A., and Dasgupta, S. (2024).
\newblock Abstract po3-16-05: Mitochondrial dna mutation detection in tumors
  and circulating extracellular vesicles of triple negative breast cancer
  patients for biomarker development.
\newblock {\em Cancer Research}, 84(9\_Supplement):PO3--16.

\bibitem[Vikramdeo et~al., 2023]{vikramdeo2023profiling}
Vikramdeo, K.~S., Anand, S., Sudan, S.~K., Pramanik, P., Singh, S., Godwin,
  A.~K., Singh, A.~P., and Dasgupta, S. (2023).
\newblock Profiling mitochondrial dna mutations in tumors and circulating
  extracellular vesicles of triple-negative breast cancer patients for
  potential biomarker development.
\newblock {\em FASEB BioAdvances}, 5(10):412.

\bibitem[Yeung and Petrosyan, 2006]{yeung2006cooperative}
Yeung, D.~W. and Petrosyan, L.~A. (2006).
\newblock {\em Cooperative stochastic differential games}, volume~42.
\newblock Springer.

\end{thebibliography}
\end{document}